\documentclass[12pt]{article}
\usepackage[dvips]{graphicx}
\newcommand{\bJ}{\mbox{\boldmath $J$}}
\newcommand{\bomega}{\mbox{\boldmath $\omega$}}
\newcommand{\bmu}{\mbox{\boldmath $\mu$}}

\newcommand{\varDelta}{{\mit\Delta}}
\newcommand{\varOmega}{{\mit\Omega}}
\newcommand{\braket}[1]{{\langle #1 \rangle}}
\newcommand{\gtsim}{\mathrel{\hbox{\raise0.2ex
    \hbox{$>$}\kern-0.75em\raise-0.9ex\hbox{$\sim$}}}}
\newcommand{\ltsim}{\mathrel{\hbox{\raise0.2ex
    \hbox{$<$}\kern-0.75em\raise-0.9ex\hbox{$\sim$}}}}

\begin{document}

\vspace*{-12mm}
\begin{center}
\begin{minipage}{18.0cm}
\begin{center}
  {\Large \bf Calculation of strongly-coupled rotational bands \\
        in terms of the tilted axis cranking model }
\end{center}
\end{minipage}

\vspace*{8mm}

  {\large Shin-Ichi Ohtsubo and Yoshifumi R. Shimizu } \\
\vspace*{2mm}
 {\it \large Department of Physics, Kyushu University,
                                       Fukuoka 812-8581, Japan}
\end{center}
\vspace*{3mm}
%%%%%%%%%%%%%%%%%%%%
%     Abstract     %
%%%%%%%%%%%%%%%%%%%%
\begin{center}
\leftline{\hspace{5mm} \large \bf Abstract }
\vspace{3mm}
\begin{minipage}{16.8cm}
\small

     Recently observed strongly-coupled rotational bands associated with
the $\nu [505]\frac{11}{2}^-$ quasiparticle state are studied by means of
a microscopic tilted axis cranking (TAC) model.
The results of calculation for the routhians
and the $B(M1)/B(E2)$ ratios are
investigated in the light of other existing models,
i.e. the strong-coupling model and the conventional cranking model.
It is demonstrated that only the TAC model can successfully
reproduce these two observables at the same time.
The reason of the success is clarified by making connections
between these models.

\end{minipage}
\end{center}

\vspace*{3mm}

{\it PACS} number(s): 21.60.-n

{\it keywords}: High-$K$ rotational band;
  Tilted axis cranking; Cranked shell model; $B(M1)/B(E2)$

\vspace*{3mm}
%%%%%%%%%%%%%%%%%%%%%%%%%%%%%%%%
%         Introduction         %
%%%%%%%%%%%%%%%%%%%%%%%%%%%%%%%%
\section{ Introduction }
\label{sect:Intro}

     Bohr-Mottelson's strong-coupling model~\cite{ref:BMt75}
is the first pioneering work
which is capable to describe interplay between the collective
and the single-particle rotational motions in well-deformed nuclei.
It successfully reproduce both
the energy spectra and the electromagnetic transition properties
of rotational bands in odd nuclei.
It is, however, of phenomenological nature since essential
quantities like the moment of inertia and the quadrupole moment
are model-parameters and are adjusted in comparison with experimental data.
One has to combine it with a more microscopic model such as
the Nilsson single-particle model.
Another drawback of the model is that it is based on the adiabatic
assumption of the collective rotation and the application to
the high angular momentum regime is not straightforward.
The particle-rotor model~\cite{ref:BMt75,ref:RSt80} is a possible means
to extend the idea by lifting the adiabatic assumption and including
the effect of Coriolis coupling by exact diagonalization,
although it is still
semi-phenomenological in the sense that the macroscopic ``rotor'' part is
explicitly introduced.  Nowadays, fully microscopic approaches
are available, for example, the variation after full angular momentum
projection base on the generalized mean-field~\cite{ref:SGF87} and
the projected shell model~\cite{ref:Ha95}.
However, they are very complicated and lose simplicity of the model.

     On the other hand, by taking into account the effect
of rotation unperturbatively, the mean-field theory has been
extended in the rotating frame:  The cranking model or
the Cranked Shell Model (CSM)~\cite{ref:BF79},
which is simple and yet microscopic, has been
successfully applied to understand various
high-spin phenomena~\cite{ref:Ga83}
such as backbending of moment of inertia.
Recently, this cranking model has been further extended in such a way
to include the tilting degrees of freedom of rotation-axis
relative to the deformed shape;
the Tilted Axis Cranking (TAC) model~\cite{ref:Fra93,ref:Fra00,ref:Fra01}
(the conventional cranking is called the
Principal Axis Cranking (PAC) model, instead),
which gives nice interpretations of new types of
nuclear rotational motions,
e.g. the shears bands~\cite{ref:Clark00} and
the chiral bands~\cite{ref:Fra01,ref:FM96}.
Although these cranking models treat the rotational motion
in a semiclassical manner, its simplicity allows us to
have a clear physical picture of various collective rotational motions,
which are actually a result of the complex nuclear many-body problem.

     The purpose of the present paper is two folds: The first is to give
a clear relationship between the strong-coupling model and the TAC model.
The second is to apply the TAC model to the strongly-coupled
high-$K$ one-quasiparticle bands, which have been recently measured
at JAERI up to high-spin states ($I \ltsim 22\hbar$) in several
nuclei in the light rare earth region.
In this way, it will be demonstrated that the TAC method is a powerful
tool to investigate the high-spin phenomena,
including not only such new types of rotational bands
but also well-known typical rotational bands,
and thus gives a good description of the nuclear rotational motions
from low- to high-spin states.

     In Section 2, the relation between the TAC model and
the strong-coupling model is studied.
The results of the TAC model are presented in Section 3
in comparison with the experimental data.
Section 4 is devoted to the conclusion.

\vspace{3mm}
%%%%%%%%%%%%%%%%%%%%%%%%%%%%%%%
%         Formulation         %
%%%%%%%%%%%%%%%%%%%%%%%%%%%%%%%
\section{ Relation between strong-coupling model and TAC model }
\label{sect:Form}

     Since the strong-coupling model is well-known~\cite{ref:BMt75},
we only recapitulate the final expressions necessary for
the following discussion.  Our main object of
study is rotational band of odd well-deformed axially symmetric nuclei
with a valence nucleon occupying a high-$K$ orbit.  Therefore, we use
the simplest expressions for the axially symmetric case
without the decoupling terms, and higher order Coriolis coupling
is neglected completely:
The energy spectrum is given by\footnote{
  We use $\hbar=1$ unit throughout in this paper.}
\begin{eqnarray}
  E_K ( I ) \, = \, \frac{1}{ 2 \mathcal{J} }
            \left[ I( I+1 ) - K^{2} \right] \, + \,
     E_K^{0} - \frac{K}{ 2 \mathcal{J} }\quad (I \ge K),
  \label{eq:STene}
\end{eqnarray}
and the $E2$ and $M1$ transition probabilities are given by
\begin{eqnarray}
  B(E2;I \to I-2) &=& \frac{5}{16\pi}\,{Q_{0}^{2}}\,
    {\langle IK20 | I-2K \rangle^{2}},
  \label{eq:STBE2} \\
  B(M1;I \to I-1) &=& \frac{3}{4\pi}\,(g_{K}-g_{R})^{2} K^{2}\,
    {\langle IK10 | I-1K \rangle^{2}},
  \label{eq:STBM1}
\end{eqnarray}
respectively.   Here the introduced parameters are the moment of inertia,
$\mathcal{J}$, the band head energy, $E_K^0$, the quadrupole moment, $Q_0$,
and the intrinsic and rotational $g$-factors, $g_K$ and $g_R$.

     Details of the microscopic TAC model
is given in Ref.~\cite{ref:Fra00,ref:Fra01}.  We follow these
references and briefly discuss about what is necessary
in the following discussion.
As for the starting microscopic hamiltonian, we take the Nilsson
single-particle potential~\cite{ref:NIL95} with
the $ls$ and $l^2$ parameters taken from Ref.~\cite{ref:Ben85},
and the monopole pairing residual interaction.
Thus, using the axial symmetry of the potential,
the single-particle (two-dimensional) TAC hamiltonian is
\begin{eqnarray}
  \hat{h}^{\prime} =  \hat{h}_{\rm Nils} \, - \, \lambda \hat{N}
  - \, \varDelta ( \hat{P}^{\dagger} + \hat{P} )
  - \, (\omega_{x} \hat{J}_{x} + \omega_{z} \hat{J}_{z}), 
  \label{eq:TACham}
\end{eqnarray}
where $\lambda$ and $\varDelta$ are the chemical potential and
the pairing gap, $\hat{N}$, $\hat{P}^{\dagger}$ and
$(\hat{J}_x, \hat{J_z})$ are
the number operator, the monopole pairing operator, and
($x,z$)-components of the angular momentum vector operator, respectively.
The characteristic feature of the TAC model is the inclusion
of the tilting degree of freedom of rotation-axis
with respect to the principal axes ($x,y,z$)
of the deformed shape, which is represented in
the cranking term $-\bomega \hat{\bJ}=
-(\omega_{x} \hat{J}_{x} + \omega_{z} \hat{J}_{z})$.
The (two-dimensional) components of the rotational frequency vector
${\bomega}$ are given by
the rotational frequency $\omega$ and
the tilting angle of the frequency vector $\theta_{\omega}$;
\begin{eqnarray}
  \cases{ \omega_{x} = \omega \sin \theta_{\omega} , \cr
          \omega_{z} = \omega \cos \theta_{\omega}.   }
 \label{eq:TACfreq}
\end{eqnarray}
The expectation value of the angular momentum operators is related
to the angular momentum quantum number $I$ by
\begin{eqnarray}
  J \equiv \sqrt{ \, \langle \hat{J}_{x} \rangle^{2}
                    + \, \langle \hat{J}_{z} \rangle^{2} }
    = I \, + \, \frac{1}{2}.
  \label{eq:SPINSHIFT}
\end{eqnarray}
Here the shift $+\frac{1}{2}$ is due to the quantum correction
of the rotational motion.  Note that the rotational frequency
$\omega$ in Eq.~(\ref{eq:TACfreq}) is a conjugate variable of $J$, namely,
\begin{eqnarray}
   \omega = \frac{d E}{d J} = \frac{d E}{d I},
  \label{eq:TACrot}
\end{eqnarray}
and the total routhian is defined by $E^\prime = E - \omega J$.
The tilting angle of the angular momentum vector
is similarly defined as
\begin{eqnarray}
  \theta_{J}   =  \tan^{-1}\left( \,
   { \langle  \hat{J}_{x} \rangle }/
   { \langle \hat{J}_{z}  \rangle }   \right)
   =\cos^{-1}\left({ \langle \hat{J}_{z}  \rangle }/J \right).
  \label{eq:AngJ}
\end{eqnarray}
This angle is determined at each rotational frequency
by the selfconsistency condition:  The routhian
in the rotating frame is minimum (stationary) with respect
to the angle of the frequency vector, which is equivalent to the condition
that two tilting angles coincide:
\begin{eqnarray}
  \left. \frac{ \partial E^{\prime} }{\partial \theta_{\omega} }
  \right|_{\omega} \, = \, 0 \quad\iff\quad
  \theta_{\omega} \, = \, {\theta}_{J}.
\label{eq:Vect-para}
\end{eqnarray}
Once the angle $\theta_{\omega}$, or ${\theta}_{J}$, is determined
at each rotational frequency,
the routhian or the energy of the system is calculated according to
the usual mean-field approximation as a function of
the rotational frequency $\omega$,
or of the angular momentum $I=J-1/2$, if necessary.

     In the present paper, the monopole pairing gap, $\varDelta$,
as well as the deformation parameters are kept constant,
while the chemical potential is determined by the number condition.
Then, in order to guarantee
the condition (\ref{eq:Vect-para}), one has to use the following
definition of the routhian,
\begin{eqnarray}
  E^{\prime} \, = \, \langle \hat{h}_{\rm Nils} \rangle
       \, - \, 2 \varDelta \langle \hat{P}^{\dagger} \rangle
       \, - \, \langle \bomega \hat{\bJ} \rangle ,
  \label{eq:Def-tene-chem}
\end{eqnarray}
in place of the usual definition,
\begin{eqnarray}
  E^{\prime} \, = \, \langle \hat{h}_{\rm Nils} \rangle
       \, - \, G  \langle \hat{P}^{\dagger} \rangle ^{2}
       \, - \, \langle \bomega \hat{\bJ} \rangle ,
  \label{eq:Def-tene-full}
\end{eqnarray}
where $G$ is the strength of the monopole pairing interaction, and
the pairing gap in this case is determined selfconsistently
by $\varDelta=G  \langle \hat{P}^{\dagger} \rangle$.

     The transition probabilities in the TAC model are
given by~\cite{ref:Fra00}
\begin{eqnarray}
  B(E2;I \to I-2) &=& \frac{5}{16\pi}\,\frac{3}{8}\,
    \left( \langle \hat{Q}_{0} \rangle^2\, \sin^2\theta_J \right)^2,
  \label{eq:TACBE2} \\
  B(M1;I \to I-1) &=& \frac{3}{4\pi}\,\frac{1}{2}\,
    \left( \braket{\hat{\mu}_z}\sin\theta_J
   - \braket{\hat{\mu}_x}\cos\theta_J \right)^2,
  \label{eq:TACBM1}
\end{eqnarray}
where $\hat{Q}_0=\sum_{a=1}^A (2\hat{z}^2 - \hat{x}^2 - \hat{y}^2)_a$,
is the quadrupole moment operator around the $z$-axis,
and $(\hat{\mu}_x,\hat{\mu}_z)$ is ($x,z$)-component of
the magnetic moment vector operator,
$\hat{\bmu}=\sum_{a=1}^A (g_l \hat{\mbox{\boldmath $l$}}
+ g^{\rm eff}_s \hat{\mbox{\boldmath $s$}} )_a$.
Here the tilting angle $\theta_J$ appears and the expectation values
of the operators in the body-fixed coordinate ($x,y,z$) are used.
The following shape-consistency constraints should be satisfied
at each frequency in order to define the body-fixed frame~\cite{ref:KO81}:
\begin{eqnarray}
  \langle \sum_{a=1}^A (\hat{x}\hat{y})_a \rangle \, = \,
  \langle \sum_{a=1}^A (\hat{y}\hat{z})_a \rangle \, = \,
  \langle \sum_{a=1}^A (\hat{z}\hat{x})_a \rangle \, = \, 0.
 \label{eq:SHAPE}
\end{eqnarray}
In the present paper, however, we use the fixed frame defined
at the zero frequency for simplicity.
We have checked that the ratio
$| \langle \sum_{a=1}^A (\hat{z}\hat{x})_a \rangle |
/| \langle \hat{Q}_0 \rangle |$ is less than $5 \times 10^{-3}$
at all rotational frequencies
in the present calculations (other constraints are automatically satisfied
because of the two-dimensional nature of the cranking term),
and so we believe that the breaking of the shape-consistency does no harm.

     The TAC prescription contains the conventional cranking (PAC) model
as a limiting case: Taking the $z$-axis as a symmetry-axis and
the $x$-axis as a rotation-axis, the PAC model is realized
if the selfconsistently determined tilting angle is $\theta_J=90^\circ$.
The important difference of the TAC ($\theta_J \ne 90^\circ$) and
PAC ($\theta_J=90^\circ$) models is that the signature $\alpha$,
the eigenvalue of 180$^\circ$-rotation about $x$-axis,
is a good quantum number in the PAC while the signature symmetry
is broken in the TAC.
Thus the $M1$ transitions only occur between the signature partner bands,
and the $B(M1)$ in the PAC model is calculated by~\cite{ref:HS79,ref:Mats88}
\begin{eqnarray}
  B \left( M1; (I,\alpha=-{\textstyle\frac{1}{2}}) \longleftrightarrow
  (I \mp 1,\alpha=+{\textstyle\frac{1}{2}}) \right)
  \quad\quad\quad\quad\quad\quad\quad\quad && \nonumber \\
    = \Bigl| \Bigl\langle \alpha=-{\textstyle\frac{1}{2}} \Bigl|
  \frac{1}{\sqrt{2}} \left\{
  i \mathcal{O} (M1,y) \pm \mathcal{O} (M1,z) \right\}
     \Bigr| \alpha=+{\textstyle\frac{1}{2}} \Bigr\rangle \Bigr|^{2},
\label{eq:PACBM1}
\end{eqnarray}
with the transition operator
\begin{eqnarray}
  \mathcal{O} ( M1, \nu ) \equiv \sqrt{\frac{3}{4 \pi}}
  \left(  \hat{\mu}_\nu - \tilde{g}_{\rm R} \hat{J}_\nu \right),
  \quad\quad
  \tilde{g}_{R} = \langle \hat{\mu}_{x} \rangle_0 /
      \langle \hat{J}_{x} \rangle_0,
  \label{eq:PACgfac}
\end{eqnarray}
where $\langle \,*\, \rangle_0$ means that the expectation value
is taken with respect to the zero-quasiparticle state
(the even-even core state) at each rotational frequency.
Note that the quantity $\tilde{g}_{R}$
play a similar role as the rotational $g$-factor in Eq.~(\ref{eq:STBM1}),
which comes from the effect of the rotational Nambu-Goldstone
mode of even-even core nucleus~\cite{ref:Mats88}.

     Now it is the place to consider the relation between the strong-coupling
model and the TAC model.  As is discussed in Ref.~\cite{ref:Fra00},
the TAC solution corresponding to the high-$K$ rotational band
starts to appear from the so-called ``band head frequency''
$\omega_{\rm BH}$; namely as long as $\omega < \omega_{\rm BH}$
the selfconsistent tilting angle stays at $\theta_J=0^\circ$.
In the strong coupling model, where the $z$-component of the angular momentum
(the projection of angular momentum to the symmetry axis) is $K$,
the angle between the the angular momentum vector and the symmetry axis
can be introduced through
\begin{eqnarray}
    \theta_K(I) \, \equiv \, \cos^{-1}(K/J)
	=\cos^{-1}\left( \frac{K}{\textstyle I+\frac{1}{2}} \right),
  \label{eq:Ang-semi}
\end{eqnarray}
because of Eq.~(\ref{eq:SPINSHIFT}).  Since the rotational frequency
of the strong-coupling band~(\ref{eq:STene})
is given by
$\omega_K(I)=\partial E_K/\partial I=(I+\frac{1}{2})/\mathcal{J}$,
the frequency corresponding to $\theta_K(I)=0$ is~\cite{ref:Fra00}
\begin{eqnarray}
    \omega_{\rm BH} =  \frac{ K }{ \mathcal{J} }
	\,\,\,(\hbox{strong-coupling model}).
  \label{eq:freq-BH}
\end{eqnarray}
This frequency may be compared with the band head frequency $\omega_{\rm BH}$
in the TAC model calculation.

     Next let us consider the transition probabilities.  Using the asymptotic
values ($I \gg 1$) of the Clebsch-Gordan
coefficients~\cite{ref:SN96,ref:Ohtsubo97},
\begin{eqnarray}
  \langle IK20 | I-2K \rangle & \approx &
   \sqrt{\frac{3}{8}} \left(1 - \left(\frac{K}{I-\frac{1}{2}}\right)^2\right)
   = \sqrt{\frac{3}{8}}\,\sin^2\theta_K(I-1),
   \label{eq:ASYME2} \\
  \langle IK10 | I-1K \rangle & \approx &
   \sqrt{\frac{1}{2}}\sqrt{ 1 - \left(\frac{K}{I}\right)^2 }
   = \sqrt{\frac{1}{2}}\,\sin\theta_K({\textstyle I-\frac{1}{2}}),
   \label{eq:ASYMM1}
\end{eqnarray}
it is apparent that
the expression of $B(E2)$ in the strong-coupling model~(\ref{eq:STBE2})
corresponds to the one in the TAC model~(\ref{eq:TACBE2}) if the TAC
tilting angle $\theta_J$ is identified with $\theta_K(I)$
in Eq.~(\ref{eq:Ang-semi}) at $I=I-1$, the mean value of
the initial and final angular momenta.
As for the $B(M1)$, introducing $g$-factors similar to that
in Eq.~(\ref{eq:PACgfac}),
\begin{eqnarray}
  \bar{g}_{K} = \langle \hat{\mu}_{z} \rangle /
      \langle \hat{J}_{z} \rangle,\quad\quad
  \bar{g}_{R} = \langle \hat{\mu}_{x} \rangle /
      \langle \hat{J}_{x} \rangle,
  \label{eq:TACgfac}
\end{eqnarray}
the quantity in the parenthesis in Eq.~(\ref{eq:TACBM1}) is written as
\begin{eqnarray}
    \braket{\hat{\mu}_z}\sin\theta_J
   - \braket{\hat{\mu}_x}\cos\theta_J =
   (\bar{g}_K - \bar{g}_R) \langle \hat{J}_z \rangle \, \sin\theta_J,
 \label{eq:TACmag}
\end{eqnarray}
which clearly indicates a equivalence between Eqs.~(\ref{eq:STBM1})
and~(\ref{eq:TACBM1}) if the $g$-factors are the same,
$\braket{\hat{J}_z}=K$, and $\theta_J$ are identified
with $\theta_K(I)$ at $I=I-\frac{1}{2}$, again the mean value of
the initial and final angular momenta.
The effect of finite $K$-values in the Clebsch-Gordan coefficients
has been recognized and similar correction terms
to the formula in the PAC model
have been used in Refs.~\cite{ref:Do87, ref:Osh89}
for transition probabilities.
For the high-$K$ orbitals the Coriolis coupling
is not effective even at the high-spin states,
so that the signature splitting is very small.
Then, the matrix element of $\mathcal{O}(M1,y)$ in Eq.~(\ref{eq:PACBM1})
is expected to be vanishing~\cite{ref:Mats88}, and
the three expressions of $B(M1)$,
Eqs.~(\ref{eq:STBM1}),~(\ref{eq:TACBM1}), and~(\ref{eq:PACBM1})
tends to coincide in the limit, $\theta_J=\theta_K(I) \rightarrow 90^\circ$,
if all the $g$-factors are the same and
$\braket{\hat{J}_z}_{\rm TAC}
= \langle \alpha=-{\textstyle\frac{1}{2}}| \hat{J}_z|
\alpha=+{\textstyle\frac{1}{2}} \rangle_{\rm PAC}=K$.
We will see that this trend is observed for $B(M1)$
in the actual cases studied in the next section.

\vspace{3mm}
%%%%%%%%%%%%%%%%%%%%%%%%%%%%%%
%         Discussion         %
%%%%%%%%%%%%%%%%%%%%%%%%%%%%%%
\section{ Results and discussions }
\label{sect:Diss}

     In this section, the results of simplest TAC calculations
are presented for high-$K$ one-quasiparticle bands.
It will be demonstrated that the TAC model is a powerful microscopic method
to describe the strongly-coupled rotational bands up to high-spin states.
Recently, rather systematic measurements, using
the Coulomb excitation and the in-beam $\gamma$-ray spectroscopy,
have been performed at JAERI (Japan Atomic Energy Research Institute)
for stable neutron-odd nuclei around $^{155}$Gd nucleus,
$^{153}$Sm \cite{ref:Haya00,ref:Re79},
$^{155}$Gd \cite{ref:Haya99},
$^{157}$Gd \cite{ref:Haya00-a},
$^{157}$Dy \cite{ref:Haya00-a}
and $^{159}$Dy \cite{ref:Suga01}
(see also the Nuclear Data~\cite{ref:Web-NNDC}).
In these series of experiments, both the positive-
and negative-parity yrast bands have been identified
up to high-spin states ($I \ltsim 25\hbar$).
They are $\varDelta I = 2$ aligned (weakly-coupled) rotational bands
with signature $\alpha=\pm\frac{1}{2}$, and positive-parity (Band 1)
and the negative-parity (Band 2) bands are supposed to
be the one-quasiparticle bands associated with the Nilsson orbitals
$\nu$[651]$\frac{3}{2}^{+}$ and $\nu$[521]$\frac{3}{2}^{-}$, respectively.
In addition to these observations, a $\varDelta I = 1$ strongly-coupled
rotational sequence with negative-parity has also been measured
up to high-spin states ($I \ltsim 22\hbar$),
which is supposed to be based on
the high-$K$ $\nu$[505]$\frac{11}{2}^{-}$ configuration.

     We have applied the (two-dimensional) TAC model in \S\ref{sect:Form}
to these measured rotational bands.  The results for the Band 1 and 2
give the selfconsistent tilting angle $\theta_J = 90^\circ$
even at lowest frequencies, and show rotationally aligned (decoupled) bands
with signature-splitting, which is large for Band 1 and moderate for Band 2;
namely, they are found to be well described in the conventional PAC model.
Only the high-$K$ (or high-$\varOmega$) $\nu h_{11/2}$ band
has the selfconsistent tilting
angle, $0 < \theta_J < 90^\circ$, in the observed rotational frequencies,
corresponding to the observed signature-splitting that is negligibly small.
In the following we mainly concentrate on this high-$K$ band.

     First, let us explain the details of the calculation:
Assuming axially symmetric deformation ($\gamma=0^{\circ}$),
selfconsistent values of the Nilsson deformation parameters
($\epsilon_2,\epsilon_4$) of the ground state
have been searched by using the Nilsson-Strutinsky method
at zero rotational frequency,
where the pairing correlations are included by using the smooth pair-gap method
with $\widetilde{\varDelta}=12/\sqrt{A}$ MeV.
The result, however, is found to give systematically smaller deformations
compared to those deduced from the measured $B(E2)$ values
near the ground state.
Therefore we have multiplied calculated values of $\epsilon_2$
by a factor 1.1 (keeping calculated $\epsilon_4$ values)
in all nuclei in order to reduce the discrepancy.
The deformation parameters thus determined
are listed in Table~\ref{table:para-def}.  These nuclei have rather stable
deformation in the rotational frequency range under consideration,
and therefore we have used fixed deformation parameters in all
the following calculations.  The pairing correlation is also an important
factor to understand the properties of rotational bands.
In the test calculation, we have used the pairing force parameter
which gives the values of pairing gap
corresponding to the measured third order even-odd mass
differences~\cite{ref:BMt75} (the mean value
of neighbouring even-even nuclei for neutron),
and performed the selfconsistent pairing calculation
(c.f.~Eq.~(\ref{eq:Def-tene-full})), where the blocking effect
is properly taken into account for neutrons.
However, we have found that the moment of inertia is too small
at low frequency and the pairing reduction at high frequency
is too large.  These trends have been known in the conventional
PAC calculation for many years.
Therefore, in order to compare the calculated results with experimental data,
we have used the constant-pairing-gap calculations
(c.f.~Eq.~(\ref{eq:Def-tene-chem})) with neutron and proton pairing gaps,
$\varDelta_{\nu}$ and $\varDelta_{\pi}$, which are obtained
by multiplying the factor 0.7 and 0.9, respectively,
to the measured even-odd mass differences.
Here the stronger reduction of the neutron gap is due to the fact
that the selfconsistent blocking calculations for neutron one-quasiparticle
states give about 80\% reduction on average.
Further reduction in both the neutron and proton gaps is
to make the moment of inertia larger.
The monopole pairing gap parameters thus determined and used
in the following calculations are listed in Table~\ref{table:para-def}.
The dependence of the final result on these pairing gap parameters
will be discussed in the end of this section.

\begin{figure}
\begin{center}
\includegraphics[height=14.8cm,keepaspectratio,clip]{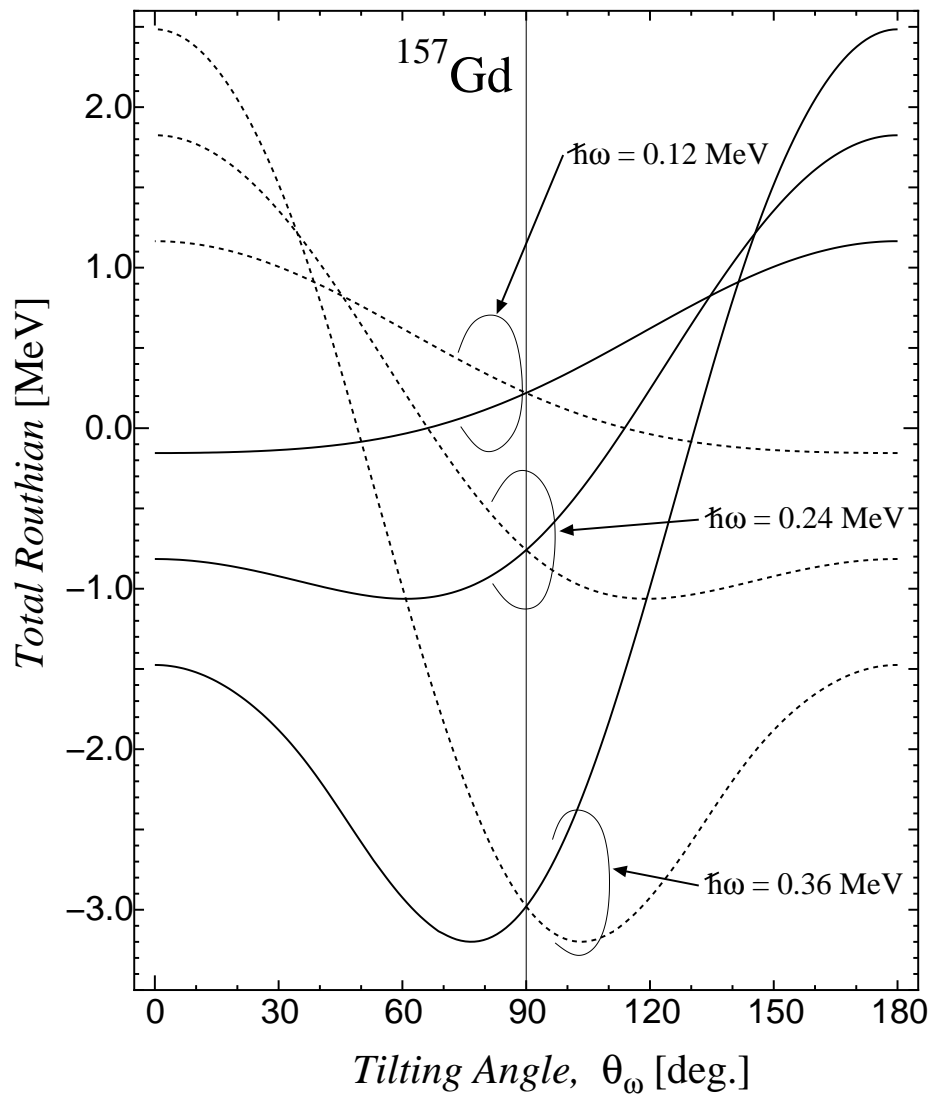}
\caption{
 The total routhians calculated for the $\nu[505]11/2^{-}$ configuration
 (high-$K$ band) in $^{157}$Gd are shown
 as functions of the tilting angle $\theta_\omega$.
 Three pairs of solid and dashed curves are the ones
 at the rotational frequency, $\hbar \omega = 0.12$, 0.24 and $0.36$ MeV.
 Two curves denoted by the solid and dotted line
 are connected diabatically starting from the $\theta_\omega=0^\circ$ and
 $\theta_\omega=180^\circ$, respectively.
}
\label{fig:Gd157-surf}
\end{center}
\end{figure}

     In Fig.~\ref{fig:Gd157-surf}, the total routhian surfaces of
the $\nu[505]\frac{11}{2}^{-}$ bands in $^{157}$Gd
at rotational frequencies $\hbar \omega = 0.12, 0.24$ and $0.36$ MeV
are shown as functions of the tilting angle $\theta_{\omega}$.
At $\theta_{\omega} = 90^{\circ}$, the TAC scheme reduces to the PAC scheme
and the calculated states are classified by
the signature quantum numbers $\alpha = \pm \frac{1}{2}$.
It is, however, noticed that the signature classification is
not appropriate for the high-$K$ orbitals, for which the $K$-quantum number
along the $z$-axis (symmetry axis) is conserved in a very good approximation
and the signature splitting is negligible.
In the figure, two surfaces shown by the full and dotted curves
have been calculated diabatically starting from states at
$\theta_{\omega} < 90^{\circ}$ and $\theta_{\omega} > 90^{\circ}$,
respectively, and the two minima correspond to the TAC states
with approximately good $K$-quantum number $+K>0$ and $-K<0$.
From the symmetry with respect to the transformation
$\theta_\omega \leftrightarrow 180^\circ - \theta_\omega$,
they are symmetric with respect to $\theta_{\omega} = 90^{\circ}$.
Note, however, that this diabatic tracing is of approximate nature
valid only for the high-$K$ orbitals, for which the Coriolis coupling between
the conjugate states with $\pm K$ are negligible
in the shown frequency range;
if the same routhian surface is considered
for the low-$K$ alignable orbits, the signature splitting at
$\theta_{\omega} = 90^{\circ}$ is sizable and the two adiabatic surfaces are
separated into lower- and higher-energy surfaces (c.f.~Ref.~\cite{ref:Fra00}).
At $\hbar\omega = 0.12$ MeV, the selfconsistent solution has
$\theta_\omega = 0^\circ$ or $180^\circ$,
while, at $\hbar\omega = 0.24$ $(0.36)$ MeV,
the selfconsistent tilting angle becomes
$\theta_{\omega} \approx 60^\circ$ $(80^\circ)$
or $120^\circ$ $(100^\circ)$.
Namely, the selfconsistent TAC solutions stay energetically favoured
compared to the conventional PAC solutions with $\theta_\omega=90^\circ$
by about $300-400$ keV in this case.

\begin{figure}
\begin{center}
\includegraphics[height=18.8cm,keepaspectratio,clip]{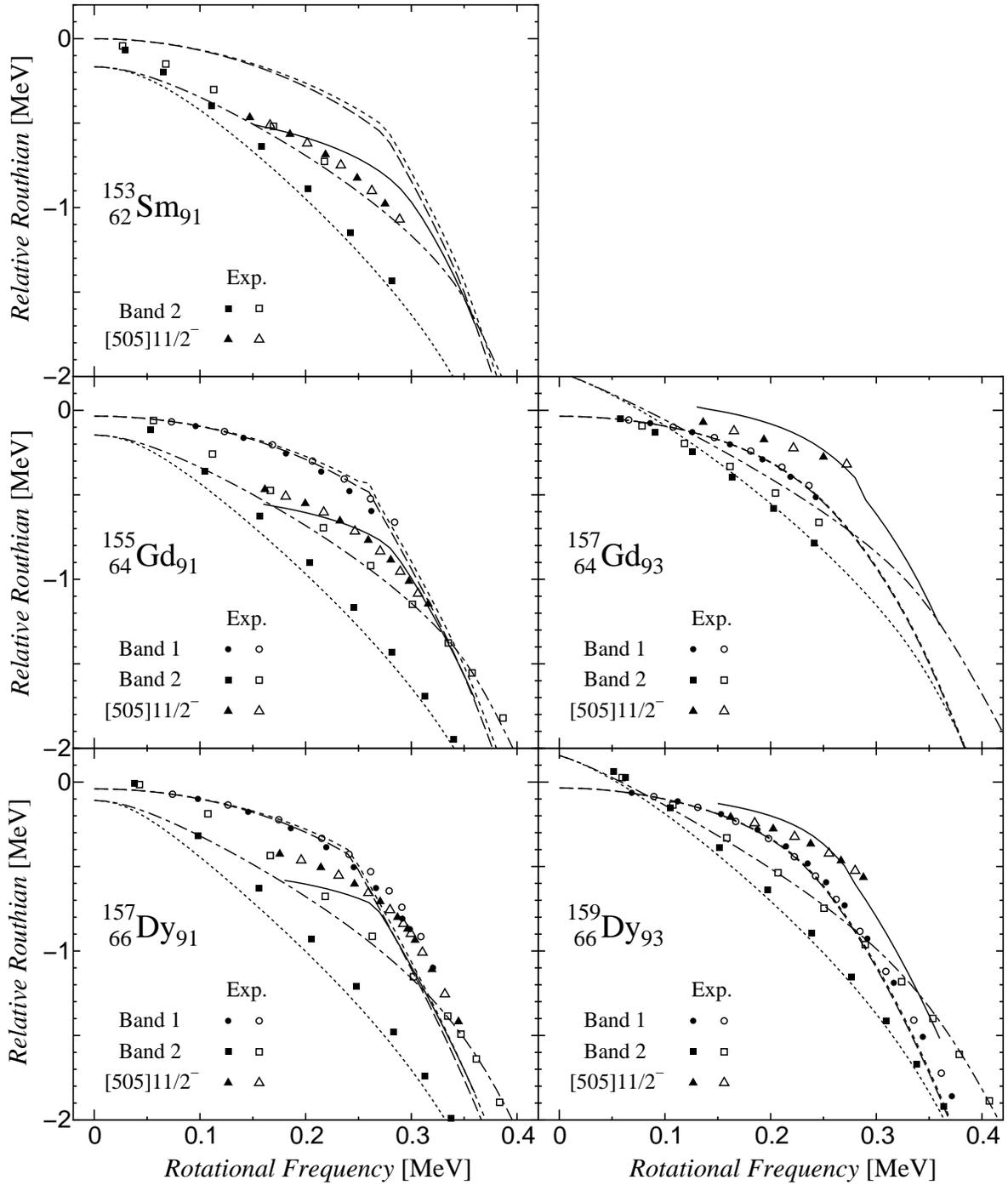}
\caption{
 Comparison of the calculated and experimental routhians
 as functions of the rotational frequency in
 $^{153}$Sm, $^{155,157}$Gd and $^{157,159}$Dy nuclei.
 The long-dashed and short-dashed lines denote calculated routhians
 with signature $\alpha = +1/2$ and $-1/2$
 for the positive-parity yrast band (Band 1),
 the dotted and dot-dashed lines denote calculated routhians
 with signature $\alpha = +1/2$ and $-1/2$
 for the negative-parity yrast band (Band 2), and
 the full line denotes the calculated routhian for the high-$K$,
 $\nu [505]\frac{11}{2}^-$ band.
 The filled and open circles denote experimental routhians
 with signature $\alpha = +1/2$ and $-1/2$ for Band 1,
 the filled and open squares denote experimental routhians
 with signature $\alpha = +1/2$ and $-1/2$ for Band 2,
 and the filled and open triangles denote experimental routhians
 with signature $\alpha = +1/2$ and $-1/2$ for the high-$K$ band.
 The ``rigid-body'' reference routhian, $-\frac{1}{2}\mathcal{J}_0\omega^2$,
 is subtraced.  Here the values of $\mathcal{J}_0$ used
 for the experimental and calculated routhians are different
 and listed in Table~\ref{table:para-def}.
}
  \label{fig:Tene-all}
\end{center}
\end{figure}

     Now let us discuss the calculated routhians in comparison with
experimental data, which are summarized in Fig.~\ref{fig:Tene-all}.
In the figure the relative routhians of Band 1, Band 2, and
the $\nu[505]\frac{11}{2}^{-}$ band are shown
as functions of the rotational frequency
in each nucleus (Band 1 was not observed in $^{153}$Sm);
the origin of the calculated routhians are chosen mainly
to reproduce Band 1.  Note that the frequency $\omega$ is defined
in Eq.~(\ref{eq:TACrot}), which is different from the conventional
PAC scheme, see discussion below.
We take a simple reference band with a constant
moment of inertia $\mathcal{J}_0$ (``rigid-body reference'' for each nucleus),
which roughly corresponds to the ground state band
in the neighbouring even-even nuclei.
Here we have to use a different
$\mathcal{J}_0$ value for calculated routhians:
This is because our calculated moment of inertia is still smaller,
even though somewhat reduced pairing gaps are used.
The $\mathcal{J}_0$ values used for calculated routhians
are 10\% smaller than those for experimental routhians in all nuclei,
which gives overall agreements between them.
The actual values of $\mathcal{J}_0$ used to make the figure are
listed in Table~\ref{table:para-def}.
This factor is consistent with the known fact that
the moment of inertia is generally underestimated
by about 20--30\% if only the monopole pairing interaction,
which reproduces the experimental even-odd mass differences,
is properly included~\cite{ref:NIL61}.   One may be able to avoid
such a phenomenological adjustment if the quadrupole pairing interaction
is included (see, e.g.~\cite{ref:YRS01}).
As already mentioned the selfconsistent tilting angle
is $\theta_{J} = 90^\circ$ in Band 1 and 2, so that they are
in fact the PAC bands and have appreciable signature splittings.
In contrast, the $\nu[505]\frac{11}{2}^{-}$ band is strongly-coupled band
and exhibits no signature splitting in the observed frequency range.
In the $N = 91$ isotones, the $\nu h_{11/2}$ band is located
between Band 1 and Band 2, while, in the $N = 93$ isotones,
it is above the Band 1.
The calculation successfully reproduces this feature.
In the $N=91$ isotones, the effect of the $i_{13/2}$ two-neutron alignment
is seen in the calculated routhians of Band 2 and the $\nu h_{11/2}$ band,
but no clear indication is observed in the experimental routhian.
Therefore it is suggested that the alignment is delayed or
the interband interaction is larger than the calculated one.
The routhian of Band 2 is slightly lower in the $N=91$ isotones,
while it is slightly higher in $^{157}$Gd.
Although there are some discrepancies we suppose that the overall
agreement is satisfactory considering that the model parameters are
not adjusted in each nucleus.

\begin{figure}
\begin{center}
\includegraphics[height=18.8cm,keepaspectratio,clip]{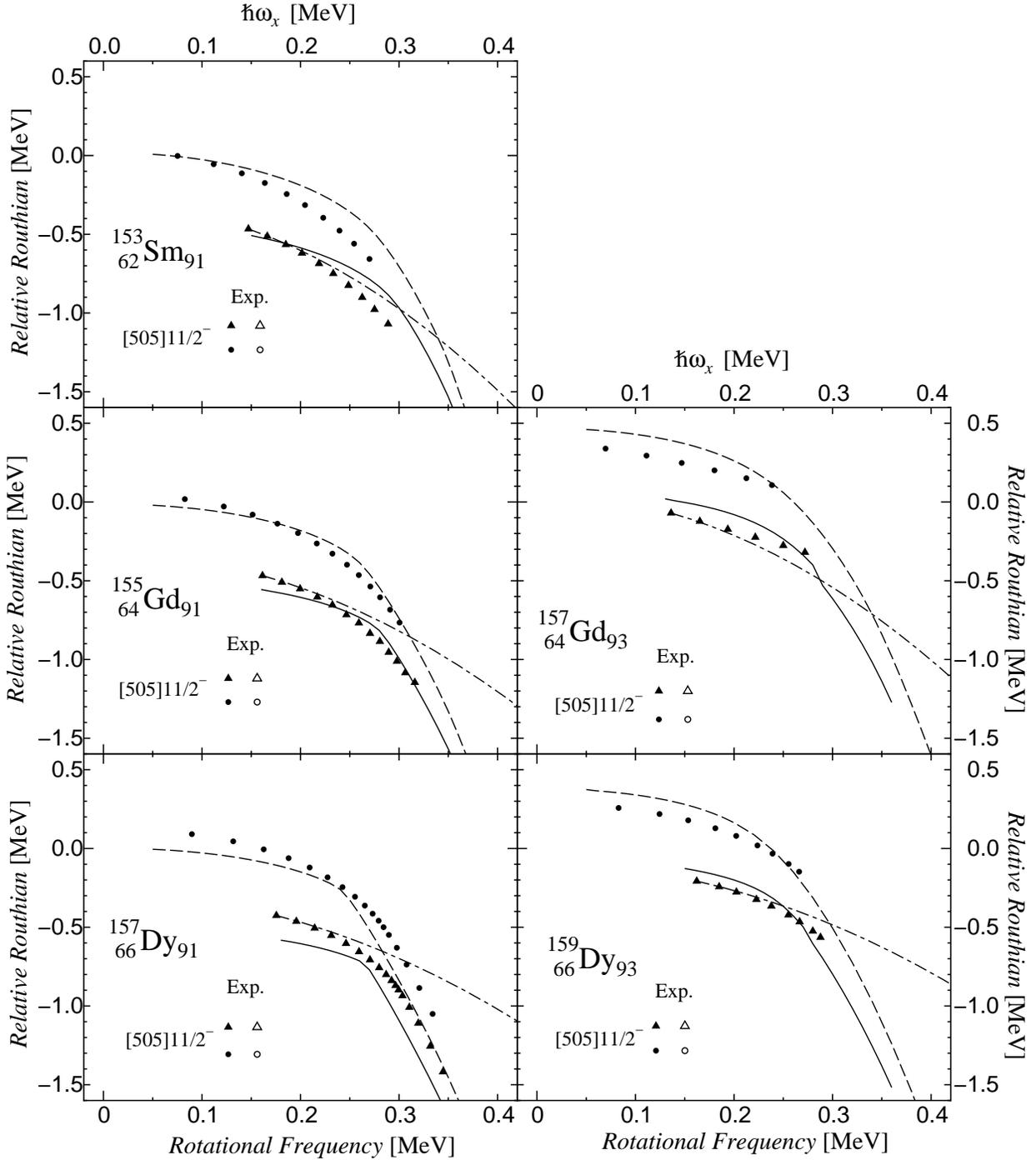}
\caption{
 Comparison of the calculated (lines) and experimental (symbols) routhians
 for the $\nu [505]\frac{11}{2}^-$ band.
 The full lines and the triangles denote the TAC scheme routhians,
 $E^\prime(\omega)$, c.f. Eq.~(\ref{eq:TACrot}), while
 the dashed lines and the circles denote the PAC scheme routhians,
 $E_x^\prime(\omega_x)$, c.f. Eq.~(\ref{eq:PACrot}).
 The dot-dashed lines also denote the calculated TAC scheme routhians,
 but using the strong-coupling expression~(\ref{eq:STrouth}),
 where the used parameters are $K =11/2$, $\mathcal{J}$ listed
 in Table~\ref{table:para-moment1}, and $E_K^{0\prime}$
 adjusted to fit the data at the lowest frequency for each nucleus.
}
 \label{fig:Ptac-all}
\end{center}
\end{figure}

    As it is discussed in Ref.~\cite{ref:Fra00},
strongly-coupled high-$K$ bands may possibly be treated in the PAC model,
i.e. in the conventional cranked shell model,
if the constancy of $K$ quantum number is good.
It should, however, be noted that the rotational frequency in the PAC model
is ``the collective rotational frequency''
(the $x$-component of the angular frequency vector) in contrast to
Eq.~(\ref{eq:TACrot}) in the TAC treatment;
\begin{eqnarray}
  \label{eq:freq-x}
  \omega_{x} =  \frac{d E}{d I_{x}} \quad{\rm with}\quad
  I_{x} ( I ) \equiv  \sqrt{ (I+{\textstyle\frac{1}{2}})^2 - K^2 },
 \label{eq:PACrot}
\end{eqnarray}
where the $z$-component of angular momentum is assumed to be
constant $K$.  Then, $\omega_x/\omega=I_x/J$, which leads
again $\theta_\omega=\theta_J$.  Correspondingly, equivalent routhians
in the PAC scheme are those in ``the collectively rotating frame''
defined by $E^\prime_x=E-\omega_x I_x$.  Thus the two states
with the same frequency in the TAC and PAC schemes have
quite different angular momenta from each other if $K$ is large.
The two routhians corresponding to $\omega$ and $\omega_x$
for the simple strong-coupling spectra~(\ref{eq:STene}) are
\begin{eqnarray}
 {\rm strong\mbox{-}coupling\,\,case:}\quad
 \cases{\displaystyle
 E^\prime(\omega)= -\frac{1}{2}\mathcal{J}\omega^2
   -\frac{K^2}{2\mathcal{J}} \, + \,E_K^{0\prime}, \cr
  \displaystyle
  E_x^\prime(\omega_x) = -\frac{1}{2}\mathcal{J}\omega_x^2
  \,+ \,E_K^{0\prime},   }
 \label{eq:STrouth}
\end{eqnarray}
where $E_K^{0\prime}=E_K^0 -(K+\frac{1}{4})/{2\mathcal{J}}$.
Thus, if two routhians,
$(\omega,E^\prime)$ and $(\omega_x,E_x^\prime)$, are plotted
in the same figure, the former is shifted by $-K^2/{2\mathcal{J}}$
relative to the latter.  The energy difference between
the TAC and PAC ($90^\circ$ tilting angle) states
in Fig.~\ref{fig:Gd157-surf} is about $300-400$ keV,
and can be interpreted in this way.
These TAC and PAC treatments for
the $\nu[505]\frac{11}{2}^{-}$ band are compared
in Fig.~\ref{fig:Ptac-all}; the solid line and triangles are
the calculated and experimental routhians in the TAC scheme, and
the dashed line and circles are
the calculated and experimental routhians in the PAC scheme.
The TAC scheme routhian for the strong-coupling expression~(\ref{eq:STrouth})
calculated by employing the moment of inertia which fits the experimental
spectra (see Table~\ref{table:para-moment1}) is also included
in the figure as the dot-dashed line.
As is stressed in Ref.~\cite{ref:Fra00} the two schemes are not strictly
equivalent because the assumption of the constant $K=\braket{J_z}$
is not valid in the selfconsistent TAC calculations, thus
the PAC scheme is only an approximation of the TAC scheme.
In fact, as is clearly seen in Fig.~\ref{fig:Ptac-all},
the calculated as well as the experimental routhians
in the two schemes are not strictly parallel, especially at high-spin
region; the distance between them gets smaller at higher frequency,
indicating that $\braket{J_z}$ reduces sizably.
However, the degree of agreement between the calculations and data
are similar in the two schemes.\footnote{
 It was discussed in Ref.~\cite{ref:Haya99} that the TAC model gives
 a better result for the routhian of Band 2 in $^{155}$Gd
 by comparing the experimental data
 with a simple CSM calculation, where the chemical potential was fixed
 by neighbouring $^{156}$Gd nucleus and used for all bands.
 It has been found that the careful treatment of the chemical potential
 in this paper makes the agreement of Band 2 much better
 even in the PAC model.}
This result suggests that the change of
$\braket{J_z}$ is not enough to show the breakdown of the PAC scheme
in this $\nu[505]\frac{11}{2}^{-}$ band.
Nevertheless, the simple strong-coupling energy~(\ref{eq:STene})
does not give a good approximation; the change of moment of inertia
as a function of the rotational frequency and the effect of
the alignments are very important to understand the high-spin spectra,
which are taken care of by both the TAC and PAC methods.

\begin{figure}
\begin{center}
\includegraphics[height=18.8cm,keepaspectratio,clip]{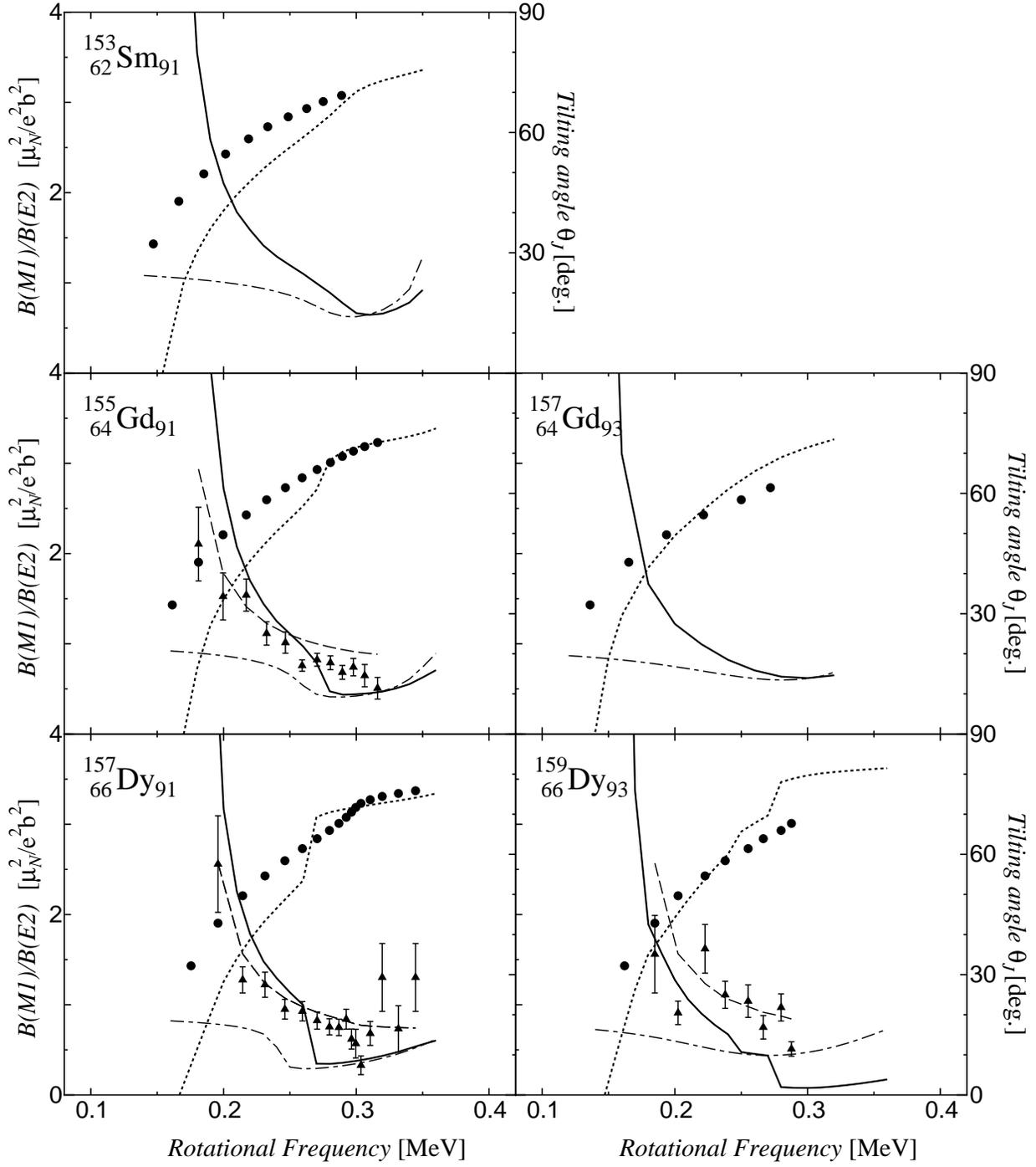}
\caption{
 Comparison of the calculated and experimental $B(M1)/B(E2)$ ratios
 for the $\nu [505]\frac{11}{2}^-$ band
 as functions of the rotational frequency in
 $^{153}$Sm, $^{155,157}$Gd and $^{157,159}$Dy nuclei.
 The solid and dot-dased lines denote the results
 by using the TAC and PAC models, respectively.  The dashed line denotes
 the result of the strong-coupling formula~(\ref{eq:BM1E2-st}),
 where the parameter $[(g_K-g_R)K/Q_0]^2$ is adjusted to fit
 the data for each nucleus.
 The triangles with error bar denote the experimetal ratios.
 In this figure, the selfconsistently determined tilting angle
 $\theta_J$ and the experimentally deduced angle~(\ref{eq:Ang-semi})
 are also included as dotted lines and circles, respectively;
 the scale for them is shown on the right hand side.
}
 \label{fig:Angt-all}
\end{center}
\end{figure}

     In addition to the energy spectra, the $B(M1)/B(E2)$ ratios
have been also measured in the JAERI experiments.
We compare the results of the calculation with experimental data
for the $\nu[505]\frac{11}{2}^{-}$ band in Fig.~\ref{fig:Angt-all}
(no $B(M1)/B(E2)$ data are available in $^{153}$Sm and $^{157}$Gd).
The selfconsistently determined the tilting angle $\theta_{J}$
and the experimentally deduced angle $\theta_K(I)$
based on Eq.~(\ref{eq:Ang-semi}) are also included in the figure.
As for the effective $g$-factors, $g_s^{\rm eff}$,
0.7 times the free values~\cite{ref:BMt69}
are used in the calculation of $B(M1)$.
The results of TAC calculation reproduce the experimental data
reasonably well, although the alignment of two-quasineutrons,
which occurs at lower frequency than the data,
makes the $B(M1)$ too small at highest frequencies.
As increasing the rotational frequency, the $B(M1)/B(E2)$ ratio
decreases, which is qualitatively understood by the strong-coupling
expressions~(\ref{eq:STBE2}) and~(\ref{eq:STBM1}), or by using
the asymptotic relations~(\ref{eq:ASYME2})
and~(\ref{eq:ASYMM1}) the ratio can be further written
as~\cite{ref:Kut95,ref:Haya99},\footnote{
  In Ref \cite{ref:Haya99}, the first line of Eq.(2) contains
  a typographical error.}
\begin{eqnarray}
  \frac{B(M1;I \to I-1)}{B(E2;I \to I-2)} & = & \frac{12}{5}
   \left[\frac{(g_{K}-g_{R})K}{Q_0}\right]^2
 \frac{\langle IK10 | I-1K  \rangle^{2}}{\langle IK20 | I-2K \rangle^{2}},
  \label{eq:BM1E2-st} \\
   {} & \approx & \frac{16}{5} \left[\frac{(g_{K}-g_{R})K}{Q_0}\right]^2
   \frac{\sin^{2} \theta_K(I-\frac{1}{2})}{\sin^{4}\theta_K(I-1)}.
 \label{eq:BM1E2-r1}
\end{eqnarray}
In the figure the ratio estimated by using the strong-coupling
expression~(\ref{eq:BM1E2-st}) is also included as the dashed line
for $^{155}$Gd and $^{157,159}$Dy, where the parameter $[(g_K-g_R)K/Q_0]^2$
is fitted to the data.
As it is clear in Eqs.(\ref{eq:BM1E2-r1}), the ratio is
quite sensitive to the geometry of the angular momentum vector
in the body-fixed frame.  In terms of the TAC scheme,
the vector of the total angular momentum is orienting toward
the $x$-axis (the axis of collective rotation) from the $z$-axis
(the symmetry axis) as increasing the rotational frequency.
From the functional form of the angle
one can estimate the band head frequency $\omega_{\rm BH}$, that is
determined by the value at which the angle starts to non-zero degree.
Using $\theta_{J}$ and $\theta_K(I)$ one obtains the calculated
and experimental $\omega_{\rm BH}$, respectively.
Furthermore, one can evaluate the moment of inertia
through Eq.~(\ref{eq:freq-BH}); the results
are listed in Table~\ref{table:para-moment1}.
As in the case of the reference moment of inertia
in Table~\ref{table:para-def}, the calculated moment of inertia
is always smaller by about 10\%.
The calculated angle $\theta_{J}$ is systematically smaller than
$\theta_K(I)$, and accordingly the calculated $B(M1)/B(E2)$ ratio
is systematically larger.  This suggests that the calculation
may be improved by including the quadrupole pairing interaction,
which increases the moment of inertia and consistently reproduces
the inertia and the even-odd mass difference~\cite{ref:YRS01}.
The dot-dashed line in Fig.~\ref{fig:Angt-all} is the calculated
$B(M1)/B(E2)$ ratio by using the PAC formula~(\ref{eq:PACBM1})
for $B(M1)$ value.  Apparently, the PAC scheme underestimates
the ratio considerably at lower frequency.  This is because the effect
of the geometry of the angular momentum vector
(or of the Clebsch-Gordan coefficient in the strong-coupling expression)
is not taken into account in the PAC model.
However, at higher frequency the TAC result approaches to the PAC one,
since the selfconsistent angle goes to $90^\circ$ (aligned to the
collective $x$-axis).  In the case of $^{159}$Dy the two calculations
do not coincide at highest frequency.  This is due to the fact that
further alignments happen in the TAC calculation and so
the two $g$-factors~(\ref{eq:PACgfac}) and~(\ref{eq:TACgfac})
deviate considerably.

\begin{figure}
\begin{center}
\includegraphics[height=18.8cm,keepaspectratio,clip]{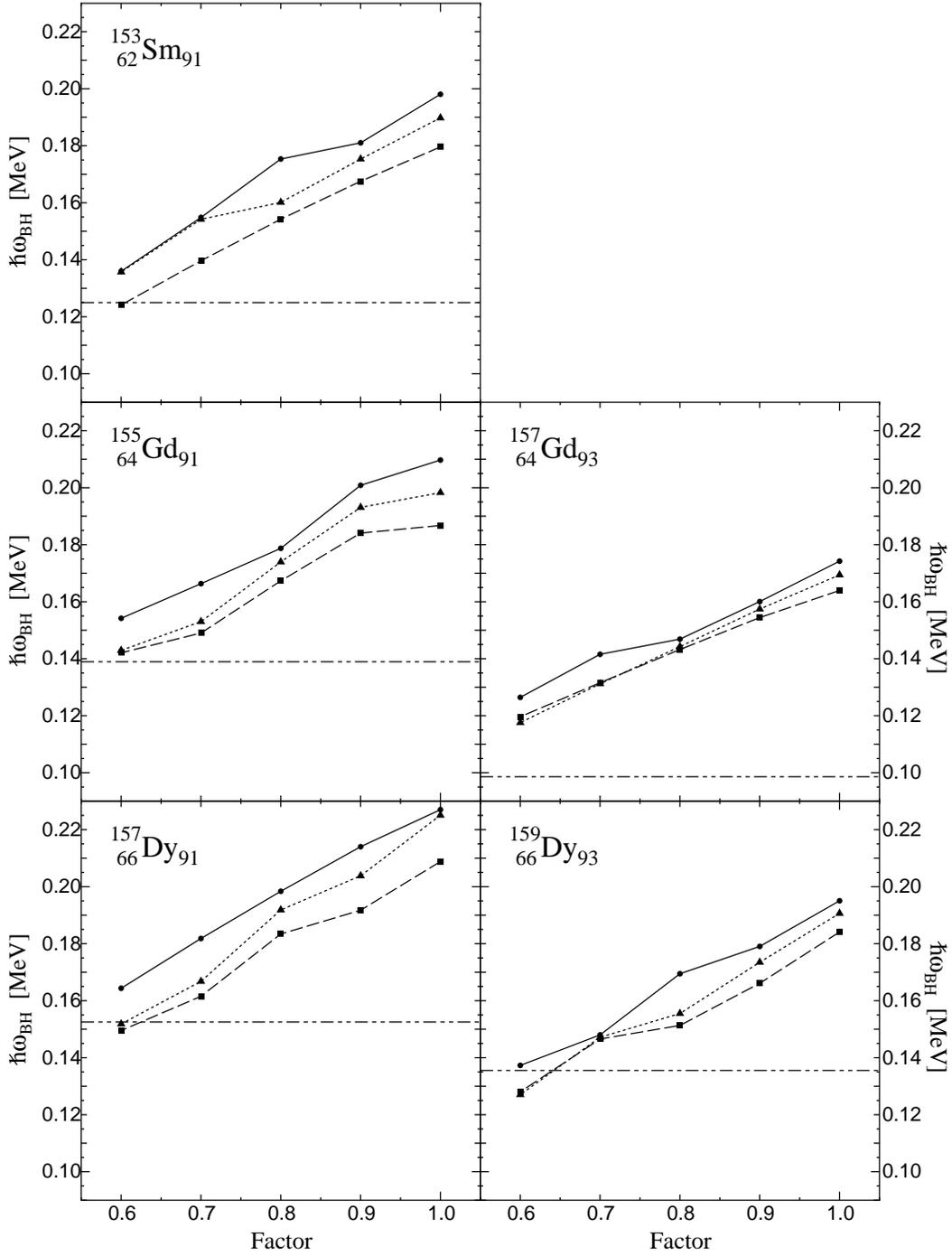}
\caption{
 Dependence of the band head frequency $\omega_{\rm BH}$
 on the neutron and proton pairing gap parameters
 in $^{153}$Sm, $^{155,157}$Gd and $^{157,159}$Dy nuclei.
 Here $\omega_{\rm BH}$ has been obtained by the TAC calculation
 where are used the neutron and proton pairing gaps
 whose values are determined by multiplying the even-odd mass differences
 by some factors.  Thus calculated $\omega_{\rm BH}$ is shown
 as a function of the factor of neutron gap.
 The solid lines with circles, dotted lines with triangles, and
 dashed lines with squares denote the results
 with the factor of proton gap being 1.0, 0.9, and 0.8, respectively.
}
 \label{fig:Angle-BH-all}
\end{center}
\end{figure}

  Finally, in order to study the dependence of the results on
the monopole pairing correlations,
the calculated band head frequencies $\omega_{\rm BH}$
are shown as functions of neutron and proton pairing gaps
in Fig.~\ref{fig:Angle-BH-all}; the qualitative feature of the results
are apparent from the relation between the $\omega_{\rm BH}$ and
the moment of inertia, Eq.~(\ref{eq:freq-BH}).
It is clear from the figure
that the reduction of the pairing gaps gives better agreement
with the experimentally deduced $\omega_{\rm BH}$.
However, if $\omega_{\rm BH}$ is fitted, the resultant pairing gaps
are too small and we cannot obtain good agreements for the routhians.
The factors 0.7 and 0.9 for neutron and proton pairing gaps
are chosen to compromise the agreements of the routhians and
the $B(M1)/B(E2)$ ratios.  In addition, we have investigated
the selfconsistent pairing calculations.  The results show that
the pairing gap reduces quite fast as increasing the rotational
frequency and the two-neutron alignments (band-crossing) occurs
at much lower frequency in contradiction to the data.
In order to avoid this early alignments, one has to use larger force
strength, which lead to the further reduction of moment of inertia.
Then, as is clear from the consideration above, the band head frequency
is considerably delayed and the agreement of the $B(M1)/B(E2)$ ratio
becomes poor.  Thus, a simultaneous reproduction of the routhians
and the $B(M1)/B(E2)$ ratio seems to be difficult if one uses
the selfconsistent pairing calculation.
This is the main reason why we have used the constant pairing gaps
with their values listed in Table~\ref{table:para-def}.

\vspace*{3mm}
%%%%%%%%%%%%%%%%%%%%%%%%%%%
%      Conclusions        %
%%%%%%%%%%%%%%%%%%%%%%%%%%%
\section{ Conclusions }

     The TAC method~\cite{ref:Fra00} has been applied to
the strongly-coupled high-$K$ band associated with
the $\nu[505]\frac{11}{2}^{-}$ Nilsson orbital,
which has been recently measured up to
high-spin states in several nuclei around $^{155}$Gd nucleus at JAERI.
The results show reasonably good overall agreement
with experimental observations for
the energy spectra (routhians) and $B(M1)/B(E2)$ ratios at the same time.
We have also investigated the same problem by using different models;
the strong-coupling model without Coriolis coupling (the lowest order
intensity relation), and the conventional
cranking model (the PAC model):\footnote{
  Of course, the strong-coupling model and the TAC as well as PAC model
  are conceptually different: In the former it is not specified
  how to calculate the essential quantities
  like the moment of inertia or the quadrupole moment, while
  such quantities are microscopically calculable in the latter models.}
The relations between these models are also clarified.
The simplest form of the strong-coupling model gives a good description
to the $B(M1)/B(E2)$ ratios if the parameters are adjusted,
but the calculated routhians deviate considerably
from the observed ones at middle or higher frequencies, which clearly
indicates the importance of the non-perturbative treatment of the Coriolis
coupling effect.  On the other hand, the conventional PAC model
is capable to describe the routhians of high-$K$ bands in almost
the same quality as the TAC model, but the agreement for
the $B(M1)/B(E2)$ ratios is poor at lower frequency,
which is due to the deficiency that a proper geometry of
the angular momentum vector with respect to the deformed-body
are not taken into account. Thus, it is concluded that
the TAC method is a simple and yet useful means to study
strongly-coupled high-$K$ rotational bands.

     It should be mentioned that the result of the present
TAC calculations is not very satisfactory.  It is necessary to use
the neutron and proton pairing gaps that are close to the even-odd
mass differences in order to achieve
a good overall agreement for all the routhians of
negative and positive yrast bands
as well as $\nu h_{11/2}$ high-$K$ band.
However, then the band head frequency for the high-$K$ band 
is overestimated, which is quite sensitive to the pairing gaps used,
and consequently the $B(M1)/B(E2)$ ratio is too large
in the low frequency region.  The sensitivity of the band head frequency
is traced back to the sensitivity of the moment of inertia
to the pairing gaps:  Note that the monopole pairing, which reproduces
the even-odd mass difference, gives too small moment of inertia.
Thus, we believe that the inclusion of the quadrupole pairing interaction,
which is known to remedy the problem of
small moment of inertia~\cite{ref:YRS01}, will improve
the agreement of both the routhians and $B(M1)/B(E2)$ ratio
between the calculations and data.
Such calculations are now in progress.

     For the one-quasiparticle bands like the one considered
in the present paper, other theoretical approaches are available,
e.g. the particle-rotor model,
or more sophisticated microscopic approach like the projected
shell model~\cite{ref:Ha95}, or the full angular momentum projection
approach~\cite{ref:SGF87}.  It is, however, increasingly more difficult
to apply them to the case of multi-quasiparticle bands.
Although the validity of the mean-field approximation should be
always carefully checked, the TAC approach is relatively easy to
apply for such more complicated rotational bands~\cite{ref:Fra00}.
We would like to extend the present study by means of the TAC method
to such cases in near feature; e.g. to the rotational bands built on
the high-$K$ multi-quasiparticle states.

\vspace*{3mm}
%%%%%%%%%%%%%%%%%%%%%%%%%%%%%%%%%%%%
%         Acknowledgments         %
%%%%%%%%%%%%%%%%%%%%%%%%%%%%%%%%%%%%
\section{ Acknowledgments }
  We are grateful to the nuclear spectroscopy group at JAERI
for providing us experimental data before publications.
Numerical calculations in this work were performed
in part by using the computer system of
Research Center for Nuclear Physics (RCNP), Osaka University,
and the Yukawa Institute Computer Facility, Kyoto University.
This work is supported in part by the Grant-in-Aid for
Scientific Research from the Japan Ministry of Education,
Culture, Sports, Science and Technology (No.~12640281).

\vspace*{3mm}
%%%%%%%%%%%%%%%%%%%%%%%%%%%%%%%
%          References         %
%%%%%%%%%%%%%%%%%%%%%%%%%%%%%%%

\clearpage

\begin{table} 
\begin{center}
\begin{tabular}{cccccccc}  \hline \hline  {} & {} & {} & {} & {} &
     \multicolumn{2}{c}{ $\mathcal{J}_{0}$ [$\hbar^{2}$/MeV] }  & {} \\
     \multicolumn{1}{c}{ \raisebox{1.5ex}[0pt]{ Nuclei            }} &
     \multicolumn{1}{c}{ \raisebox{1.5ex}[0pt]{ $\epsilon_{2}$ }} &
     \multicolumn{1}{c}{ \raisebox{1.5ex}[0pt]{ $\epsilon_{4}$ }} &
     \multicolumn{1}{c}{ \raisebox{1.5ex}[0pt]{ $\varDelta_{\nu}$[MeV]}} &
     \multicolumn{1}{c}{ \raisebox{1.5ex}[0pt]{ $\varDelta_{\pi}$[MeV]}} &
     Experiment &  Calculation \\
  \hline
   $^{153}$Sm  &  0.254  &  $-$0.0373  &  0.814  & 0.918  &  29.4  &  26.5  \\
   $^{155}$Gd  &  0.256  &  $-$0.0279  &  0.821  & 0.942  &  29.2  &  26.3  \\
   $^{157}$Gd  &  0.258  &  $-$0.0300  &  0.687  & 0.827  &  34.6  &  31.4  \\
   $^{157}$Dy  &  0.256  &  $-$0.0181  &  0.821  & 0.988  &  27.8  &  25.0  \\
   $^{159}$Dy  &  0.257  &  $-$0.0202  &  0.715  & 0.905  &  31.4  &  28.3  \\
  \hline
  \hline
\end{tabular}
\caption{
 Deformation parameter, pairing gap and
 the moment of inertia of the reference band used in the calculation.
}
\label{table:para-def}
\end{center}
\end{table}

\begin{table} 
\begin{center}
\begin{tabular}{ccccccc}  \hline \hline
  \multicolumn{2}{c}{ Nuclei } &
       $^{153}$Sm  &  $^{155}$Gd  &  $^{157}$Gd  &  $^{157}$Dy  &  $^{159}$Dy \\ \hline
    {} & Experimental &  44.2  &  40.2  &  47.8  &  37.0  &  40.1  \\
  \multicolumn{1}{c}{\raisebox{1.5ex}[0pt]{ $\mathcal{J}^{K=11/2}$ } }
       & Calculation  &  35.7  &  35.9  &  41.9  &  33.0  &  37.4  \\
  \hline
  \hline
\end{tabular}
\caption{
 Moment of inertia (in $\hbar^{2}$/{\rm MeV})
 estimated from the experimental and calculated band head frequency
 through Eq.~(\ref{eq:freq-BH}) for the $\nu[505]\frac{11}{2}^-$ band.
}
 \label{table:para-moment1}
\end{center}
\end{table} 

\clearpage

%\clearpage

\end{document}